\documentstyle[aps,pre,multicol,epsfig]{revtex}

\begin{document}

\draft

\title{Super-roughening versus intrinsic anomalous scaling of surfaces}

\author{Juan M. L\'opez $^1$, Miguel A. Rodr\'{\i}guez$^2$, 
and Rodolfo Cuerno$^3$}

\address{$^1$ Department of Mathematics, Imperial College, 
180 Queen's Gate, 
London SW7 2BZ,
United Kingdom}

\address{$^2$ Instituto de F\'{\i}sica de Cantabria, 
Consejo Superior de Investigaciones Cient\'\i ficas -- 
Universidad de Cantabria, \\
Avenida de los Castros s/n, E-39005 Santander, Spain}

\address{$^3$Departamento de Matem\'aticas y Grupo Interdisciplinar de 
Sistemas Complicados \\
Universidad Carlos III de Madrid, c/ Butarque 15, E-28911 Legan\'es, Spain}

\maketitle

\begin{abstract}

In this paper we study kinetically rough surfaces which display anomalous 
scaling in their local properties such as roughness, or 
height-height correlation function. By studying the power spectrum 
of the surface and its relation to the height-height correlation, 
we distinguish two independent causes for anomalous scaling.
One is super-roughening (global roughness
exponent larger than or equal to one), even if the spectrum behaves non 
anomalously. Another cause is what we term an
intrinsically anomalous spectrum, in
whose scaling an independent exponent exists, which induces different
scaling properties for small and large length scales (that is, the surface
is not self-affine). In this case, the surface does not need to be 
super-rough in order to display anomalous scaling. 
In both cases, we show how to extract the independent 
exponents and scaling relations from the correlation functions, and we
illustrate our analysis with two exactly solvable examples. One is the 
simplest linear equation for molecular beam epitaxy
, well known to display anomalous scaling due to 
super-roughening. The second example is a random diffusion equation, which
features anomalous scaling independent of the value of the global roughness
exponent below or above one.

\end{abstract}


\begin{multicols}{2}
\narrowtext

\section{Introduction}

The problem of the dynamics of growing surfaces
has attracted very much attention in recent years.
A large variety of discrete models and continuous
equations have been investigated numerically and analytically 
in order to explain the diversity of rough surfaces observed in 
non-equilibrium natural processes \cite{alb,revrough1,revrough2,revrough3}. 
In many cases, it is 
observed that the surface morphology displays time and space
fluctuations which have scaling properties akin to those of 
equilibrium critical phenomena.
Most of the work is thus devoted to identifying the different 
universality classes to which the 
models studied belong, which are basically determined by the
exponents characterizing the fluctuations of the interface.

We will be interested in this 
work in the dynamics of a one dimensional surface,
which will be described by the stochastic variable $h(x,t)$ giving 
the height of the surface at time $t$ above substrate
position $x$. The physical properties of the system are thus described 
by the moments of $h(x,t)$. For instance, the {\it global} width
\begin{equation}
\label{global}
W(L,t) =
\langle \langle \bigl(h(x,t) - \overline h(t) \bigr)^2 
\rangle_x ^{1/2}\rangle,
\end{equation}
is a measure of the interface fluctuations. In (\ref{global}),
$\langle...\rangle_x $ denotes spatial average over
the whole system of lateral size $L$,
$\overline h(t) \equiv \langle h(x,t) \rangle_x$ is the 
average value of the height at time $t$, and $\langle ...\rangle $ is an
average over realizations of the noise. 
Starting from a initially flat surface, 
the global width is observed to satisfy in many cases 
the dynamic scaling {\it Ansatz}
of Family-Vicsek \cite{fv},
\begin{equation}
\label{FV-globalwidth}
W(L,t) = t^{\chi/z} f(L / t^{1/z}),
\end{equation}
where the scaling function $f(u)$ behaves as
\begin{equation}
\label{FV-forf}
f(u) \sim
\left\{ \begin{array}{lcl}
     {\rm const.} & {\rm if} & u \gg 1\\
     u^{\chi}     & {\rm if} & u \ll 1
\end{array}
\right. .
\end{equation}
The roughness exponent 
$\chi $ characterizes the surface morphology in the stationary
regime, in which the correlation length $\xi(t) \sim t^{1/z}$ 
($z$ is the so called dynamic exponent) has reached a value 
larger than the system size $L$. This happens for times larger than the 
saturation time, $t \gg t_s(L)$, which scales with $L$ as $t_s(L) \sim L^z$.
The ratio $\beta= \chi/z$ is called growth exponent and
characterizes the short time behavior of the surface. 

The existence of dynamic scaling as in (\ref{FV-globalwidth}),
plus the assumption of the self-affine character
of the interface ---in the sense that there is no characteristic length scale
in the surface besides the system size, and therefore all scales obey the 
same physics--- allows one to obtain the exponents 
from the {\it local} width $w(l,t)$, 
which measures the surface height fluctuations 
over a window of size $l \ll L$. 
At times larger than 
$t_s(l) \sim l^z$ the local width is thus expected to saturate and
\begin{equation}
\label{loc-rough-exp}
w(l,t \gg l^z) \sim l^{\chi_{loc}},
\end{equation}
where the {\it local} roughness exponent $\chi_{loc}$ should
equal the exponent obtained in (\ref{FV-globalwidth},\ref{FV-forf}), 
$\chi_{loc}=\chi$.
For short times, $t \ll t_s(l)$, the local width scales
with time as $w(l,t) \sim t^\beta$, which is the short scale analog of 
the scaling behavior for the global width given 
in (\ref{FV-globalwidth}). It is an interesting fact to stress here that
the equality $\chi_{loc} = \chi$ in (\ref{loc-rough-exp})
is {\em not} 
guaranteed in general when Family-Vicsek scaling holds for the global 
width (\ref{FV-forf}), since the self-affinity of the interface is 
an additional independent condition.
  
Although the above scaling picture has been successfully applied to 
a great variety of models, which includes for instance the 
importat universality classes associated to the Edwards-Wilkinson (EW) 
\cite{ew} and Kardar-Parisi-Zhang (KPZ) \cite{kpz} equations, 
it is not always valid.   
In very recent studies of discrete models and continuum equations, 
mainly related with the growth of surfaces by molecular beam epitaxy (MBE),
it has been found that in the case of the so-called
{\it super-rough} surfaces, {\it i.e.} for surfaces with a global roughness
exponent $\chi > 1$, the usual assumption of the equivalence between
the global and local 
descriptions of the surface is not valid. 
In these systems \cite{linmbe,das1} the local width (or equivalently the 
height-height
correlation function, see below) does not saturate as expected in the standard
Family-Vicsek scaling, but crosses over to a new behavior in the
intermediate time regime $l^z \ll t \ll L^z$,
characterized by a different growth exponent $\beta_*$, 
$w(l,t \gg l^z) \sim l \; t^{\beta_*}$, where $\beta_* = \beta - 1/z$. 
It is easy to see that now the local width saturates only
when the global width does (this is, $t_s(l) = L^z$ for all $l$) and 
$w(l,t \gg L^z) \sim l L^{\chi - 1}$ in saturation, which,
according to (\ref{loc-rough-exp}),
yields a local exponent $\chi_{loc}=1$ for super-roughening.
This scaling behavior which is not encompassed by the the Family-Vicsek 
Ansatz has been termed {\em anomalous scaling} in the literature
\cite{das1,sch} and later different models have been studied in which
$\chi_{loc}$ and $\beta_*$ take values different from 1 and 
$\beta - 1/z$, respectively \cite{das2}.
 
While it is currently clearly understood that super-roughening always leads to 
anomalous scaling, in a recent paper \cite{lack} two of us 
have demonstrated that growth models in which $\chi < 1$ may also exhibit an
unconventional scaling behavior with similar scaling relations
between exponents. In that reference, for an analytically solvable growth 
model with tunable values of $\chi$, we identified anomalous dynamic scaling
with the lack of self-affinity. Also a class of discrete MBE growth models
studied numerically in Ref.\cite{das2} behaves anomalously, although the 
interfaces generated were not super-rough at all.
That is, surfaces can display anomalous scaling
{\em no matter} what their value of the global roughness exponent is, either 
$\chi > 1$ or $\chi < 1$. In the presence of anomalous scaling, not all
length scales are equivalent in the system; the scaling behavior 
is different for short (local) and large length scales, hence $\chi_{loc}$ 
and $\chi$ differ.

Our aim in the present paper is to complete the picture presented in 
Ref.\ \cite{lack}, showing that the mechanisms which lead to anomalous 
scaling behavior
can be separated into two classes, according to the behavior
of the structure factor or power spectrum of the surface, $S(k,t)$, 
to be defined below. One of the mechanisms is super-roughening, even if 
$S(k,t)$ behaves as in the Family-Vicsek case. The other mechanism 
corresponds to what we term {\em intrinsic} anomalous scaling of
the structure factor, where an independent exponent appears which measures the 
difference between the short and large length scale power laws, namely,
the difference between the local and global roughness exponents
$\chi_{loc}$ and $\chi$.
We will show how to identify the corresponding 
anomalous scaling by extracting the independent critical exponents from the 
correlation functions. As stated above, for simplicity we restric 
our further analysis to growth models in one dimension, but the conclusions 
are expected to be valid in arbitrary dimensions.
We note that, although the scaling relations to appear below for the 
super-rough and intrinsic anomalous cases exist already in the literature
\cite{sch,das1}, seemingly the independence between the occurrence of anomalous
scaling and the value (larger or smaller than one) of the global roughness
exponent $\chi$ has not been adequately understood so far. Since a clear 
picture of the different scaling behaviors which can appear is 
crucial to correctly identify them in experiments and models, we 
believe a somewhat detailed presentation is in order. It is attempted in 
Section II. The rest of the paper is organized as follows. In Section III
we illustrate the conclusions of Sec.\ II by an analysis of two growth models 
formulated in terms of stochastic equations. Even though the two models
are exactly solvable, we choose to study them numerically  in order 
to illustrate how our scaling analysis can be applied to more general
cases. The first model examined is the linear MBE equation, known to 
display anomalous scaling due to super-roughening. The second model we 
consider in Sec.\ III is the stochastic diffusion equation with random 
diffusion coefficient studied in \cite{col,lack}, which 
exemplifies a case of intrinsic anommalous scaling. Finally, we present some 
conclusions in Sec.\ IV, and discuss consequences for scaling analysis 
of rough surfaces in experiments and numerical simulations. Appendix A
contains details on the calculations discussed in Sec.\ II.
 
\section{Anomalous dynamic scaling}

In order to study a problem of kinetic roughening the 
height-height correlation function
\begin{equation}
G(l,t) = \langle \langle (h(l+x,t)-h(x,t))^{2} \rangle_x \rangle,
\label{h-h}
\end{equation} 
is frequently used.
This correlation function scales in the same
way as the square of the local width, $G(l,t) \sim w^2(l,t)$,
and provides an alternative method to determine the 
critical exponents.

However, as we will see in the following, a more transparent understanding
of the complete dynamic scaling can be obtained by studying 
the Fourier transform of the interface height in a system of
linear size $L$ (see {\em e.\ g.} \cite{alb} and references therein), 
\begin{equation}
\widehat{h}(k,t) = 
L^{-1/2} \sum_x [h(x,t) - \overline h(t)] \exp(ikx), 
\end{equation}
where the spatial average of the height has been subtracted.
In this representation, the properties of the surface can be investigated by 
calculating the structure factor or power spectrum 
\begin{equation}  
\label{S}
S(k,t) = \langle \widehat{h}(k,t) \widehat{h}(-k,t) \rangle ,
\end{equation}
which contains the same information on the system as the height-height
correlation function $G(l,t)$ defined in (\ref{h-h}), both of them being
related by \cite{inci}
\begin{eqnarray}
\label{G-from-S}
G(l,t) & = & {4 \over L} \sum_{2\pi/L\leq k \leq \pi/a} [1-\cos (kl)] S(k,t)
\\
& \propto & \int_{2\pi /L}^{\pi/a} {dk \over 2\pi} [1 - \cos(kl)] S(k,t).
\nonumber
\end{eqnarray}

\subsection{Family-Vicsek scaling}

Family-Vicsek scaling reads, when expressed in terms of the structure factor, 
\begin{equation}
\label{Sscal}
S(k,t) =k^{-(2\chi+1)} s_{FV}(kt^{1/z}),
\end{equation}
with $s_{FV}$ the following scaling function 
\begin{equation}
\label{FVforS}
s_{FV}(u) \sim  
\left\{ \begin{array}{lcl}
     {\rm const.} & {\rm if} & u \gg 1\\
     u^{2\chi + 1} & {\rm if} &  u \ll 1
     \end{array}
\right. ,
\end{equation}
Indeed, Eqs.\ (\ref{Sscal},\ref{FVforS}) can easily be seen to 
be equivalent to Eqs.(\ref{FV-globalwidth},
\ref{FV-forf}), by noting that the global width is nothing but the 
integral of $S(k,t)$, i.e.
\begin{equation}
W^2(L,t) = {1 \over L} \sum_k S(k,t) 
= \int {dk \over 2\pi} S(k,t),
\end{equation}
where the momentum integral is limited to $2\pi/L \leq k \leq \pi/a$ 
and represents a continuum aproximation to the sum over the discrete set
of modes. In a discrete growth model, $a$ is identified with the lattice
spacing. Note as well that Eq.\ (\ref{FVforS}) implies that, for $k t^{1/z}
\gg 1$, the spectrum does {\em not} depend on time, and hence at 
saturation ($t^{1/z} \gg L$), $S(k,t)$ is a pure power law independent of 
system size. 

Going back to real space, and having assumed a Family-Vicsek scaling for 
$S(k,t)$, Eq. (\ref{FVforS}), one obtains, using Eq. (\ref{G-from-S}),
\begin{eqnarray}
G(l,t) & = & \int_{\frac{2\pi}{L}}^{\frac{\pi}{a}}  
\left[1 - \cos\left(k l \right) \right]
\; \frac{s_{FV}(k t^{1/z})}{k^{2\chi +1}} \; dk \nonumber \\
& = & l^{2\chi} \int_{\frac{2\pi l }{L}}^{\frac{\pi l}{a}}  
\left[1 - \cos\left( u \right) \right]
\; \frac{s_{FV}(\rho u)}{u^{2\chi +1}} \; du , \label{int-of-G}
\end{eqnarray}
where we have introduced the ratio $\rho = t^{1/z}/l$. The limiting behaviors 
of (\ref{int-of-G}) are very simple to study when $\chi < 1$; the details
can be found in the Appendix. For instance, at saturation
$t^{1/z} \gg L$, the scaling function in the integrand of Eq.\ (\ref{int-of-G})
can be substituted for its (constant) limiting behavior at large arguments (see
(\ref{FVforS})), and one 
is left with $G(l,t) \sim l^{2 \chi}$, the limits $a \rightarrow 0$,
and $L \rightarrow \infty$ having been taken. Analogously one can arrive at
the local scaling given by
\begin{equation}
 \label{FVforG}
G(l,t) \sim  
\left\{ \begin{array}{lcl}
     t^{2\chi/z} & {\rm if} & t^{1/z} \ll l\\
     l^{2\chi}   & {\rm if} & t^{1/z} \gg l
     \end{array}
\right\} = 
l^{2\chi} \; g_{FV}(l/t^{1/z}),
\end{equation}
where the scaling function $g_{FV}(u) $ is {\em constant} for $u \ll 1$
and $g_{FV}(u) \sim u^{-2\chi}$ for $u \gg 1$.
This is the usual situation in which $\chi < 1$ and the local and global 
roughness exponents are equal ($\chi = \chi_{loc}$ and $\beta_*\equiv 0$).

\subsubsection{Super-roughening}
In the case of growth models generating super-rough surfaces ($\chi > 1$),
but with a {\em structure factor fulfilling Family-Vicsek scaling},
the integrals in (\ref{int-of-G}) are divergent in the limit
$l \ll t^{1/z}$ ($\rho \gg 1$) for $L \rightarrow \infty$, 
given the strong singularity 
at the origin of integration \cite{das1}. 
Taking the limit $l \ll t^{1/z}$ first
for fixed $a$, $L$ (see Appendix), one obtains 
\begin{equation}
G(l,t) \sim
\left\{ \begin{array}{lcl}
   l^2 \; t^{2(\chi - 1)/z} & {\rm if} & l \ll t^{1/z} \ll L  \\ 
   l^2 L^{2(\chi -1)} & {\rm if} & l \ll L \ll t^{1/z} 
\end{array}
\right. ,
\label{Gsuper}
\end{equation}
so that $\chi_{loc} =1$ and $\beta_* = \beta - 1/z$. 
In the early time regime $t^{1/z} \ll l \ll L$, 
$G(l,t) \sim t^{2\chi/z}$. 
As has been remarked in Refs.\cite{lt,rh}, the fact that $\chi_{loc}$
cannot exceed 1 for super-rough surfaces ($\chi > 1$) is a purely geometric 
property which follows from definition (\ref{h-h}).

\subsection{Intrinsic anomalous roughening}
Apart from the super-roughening case, there is another important
situation which leads to anomalous scaling. In some growth
models ---an example of which is discussed in Sec.\ III---, 
{\em the structure factor} may present an unconventional
scaling not described by Family-Vicsek (\ref{FVforS}). 
Let us consider a dynamic scaling form
for $S(k,t)$ as in Eq.(\ref{Sscal}) but with the scaling function $s_{FV}(u)$
replaced by
\begin{equation}
\label{Anom-S}
s_{A}(u) \sim  
\left\{ \begin{array}{lcl}
     u^{2\theta} & {\rm if} &  u \gg 1\\
     u^{2\chi + 1} & {\rm if} &  u \ll 1
     \end{array}
\right. ,
\end{equation}
where the label $A$ denotes intrinsic anomalous spectrum. 
Here $\theta$ is a {\em new} exponent which ``measures" the 
anomaly in the spectrum. In a system of size $L$, Eqs.\ (\ref{Sscal}), 
(\ref{Anom-S}) hold only up to the saturation time $t_s(L) = L^z$, after 
which the system size $L$ replaces the correlation length $t^{1/z}$ in all
expressions. Thus in particular, at saturation the structure factor 
depends on the size of the system as 
$S(k,t) \sim L^\theta k^{\theta-(2\chi +1)}$. 
As a consequence, the stationary spectrum shows severe finite size effects,
to the extent that it is not defined in the thermodinamic limit $L \to \infty$.

A scaling behavior such as Eqs.\ (\ref{Sscal}), (\ref{Anom-S}) 
for the structure factor does not affect the behavior of the {\em global} 
width, which preserves its Family-Vicsek form,
$W(L,t) \sim t^\beta$ for $t \ll L^z$ and 
$W(L,t \gg L^z) \sim L^\chi$. 
On the contrary, the {\em local} properties of the surface change dramatically
if $S(k,t)$ scales as in (\ref{Anom-S}). That can be
seen by computing the height-height correlation
function from Eq.\ (\ref{G-from-S}),
\begin{equation}
\label{local-A}
G(l,t) = l^{2\chi} \int_{\frac{2\pi l }{L}}^{\frac{\pi l}{a}}
\left[1 - \cos\left( u \right) \right]
\; \frac{s_{A}(\rho u)}{u^{2\chi +1}} \; du
\end{equation}
which as before (and see Appendix) gives $G(l,t) \sim t^{2\beta}$ for 
times $t \ll l^z$.
However, for intermediate times $l^z \ll t \ll L^z$ ($\rho \gg 1$) the integral 
(\ref{local-A}) picks up a non-trivial contribution from the behavior of
$s_A(u)$ at large arguments, so that 
\begin{equation}
G(l,t) \sim l^{2\chi} \rho^{2\theta} = l^{2\chi - 2\theta} t^{2\theta/z} 
\nonumber
\end{equation}   
Thus, the complete scaling of the height difference 
correlation function (or, equivalently, the square of the local width) 
can be written as
\begin{equation} 
\label{Anom-G}
G(l,t) \sim  
\left\{ \begin{array}{lcl}
     t^{2\beta} & {\rm if} & t^{1/z} \ll l \ll L \\
     l^{2\chi_{loc}} t^{2\beta_*} & {\rm if} & l \ll t^{1/z} \ll L \\
     l^{2\chi_{loc}} L^{2\theta} & {\rm if} & l \ll L \ll t^{1/z} 
     \end{array}
\right\} = l^{2\chi} g_A(l/t^{1/z}),
\end{equation}
where the local roughness exponent is $\chi_{loc} = \chi - \theta$, 
$\beta_*=\theta/z = \beta - \chi_{loc}/z$,
and the scaling function $g_A(u)$ is {\em not} constant anymore for small  
arguments, but behaves as 
\begin{equation}
g_A(u) \sim
\left\{ \begin{array}{lcl}
u^{-2(\chi - \chi_{loc})} & {\rm if} & \hspace{2mm} u \ll 1  \\
u^{-2\chi} & {\rm if} & \hspace{2mm} u \gg 1 
\end{array}
\right. .
\label{gA}
\end{equation}
The fact that
$\theta \ne 0$ in Eq. (\ref{Anom-S})
yields a local roughness exponent $\chi_{loc} \ne \chi$ and 
an anomalous growth exponent $\beta_* \ne 0$. Therefore, there are 
now {\em three}
independent exponents describing the scaling properties of the surface, 
whereas for Family-Vicsek scaling (even in the presence of super-roughening)
there are only two. Which triplet of independent
exponents one considers depends mainly on the kind of scaling properties one 
is measuring. In experiments normally it is local properties of the surface 
(local width, etc.) that are measured, and one might naturally
consider as independent exponents {\em e.g.} $\chi_{loc}$, $z$, and 
$\beta_*$. On the other hand, the focus of numerical simulations are usually
finite size effects, thus probing global properties of the surface such as the
global width or power spectrum. In the latter cases a common choice is 
{\em e.g.} $\chi$, $\theta$, and $\beta$.
We should stress that, although the scaling relations for exponents 
are exactly the same as for the super-roughening case discussed
above, the present type of anomalous scaling is due to $\theta \ne 0$
and may and does occur in growth models with $\chi  < 1$. 
The scaling behavior (\ref{Anom-G}) of the local width is formally 
equivalent to that of super-roughening (for which case
$\chi_{loc} = 1$ and $\beta_* = \beta - 1/z$), a fact which has produced some
confusion in previous works where both anomalies have been identified
in some way. 

In view of the local scaling, Eq.\ (\ref{Anom-G}), the 
structure factor can be rewritten conveniently as
\begin{equation}
\label{Scaling-S}
S(k,t) \sim  
\left\{ \begin{array}{lcl}
     t^{(2\chi +1)/z} & {\rm if} & k t^{1/z} \ll 1\\
     k^{-(2\chi_{loc}+1)} t^{2 (\chi - \chi_{loc})/ z} &
     {\rm if} & k t^{1/z} \gg 1
     \end{array}
\right. ,
\end{equation}
where we can observe two interesting facts that
characterize what we call an {\it intrinsic 
anomalous scaling}. First, $S(k,t^{1/z}  \gg k^{-1})$ 
decays as $k^{-(2\chi_{loc}+1)}$, and {\em not} following the 
$k^{-(2\chi+1)}$ law
characteristic of Family-Vicsek scaling (super-roughening case included). 
Second, there is an unconventional dependence of  
$S(k,t^{1/z} \gg k^{-1})$ on time which leads to a non stationary 
structure factor for $t^{1/z} k \gg 1$.
The combination of these two facts allows to distinguish between anomalous
scaling due to super-roughening ($\chi > 1$) and intrisic anomalous scaling.

\section{Models and numerical results}
In this section we present numerical 
simulations of two different growth equations in order to 
illustrate the theory we have just discussed in the previous 
section. First, we discuss the linear MBE equation  
as a paradigmatic example of Family-Vicsek behavior of the structure
factor, in the presence of super-roughening, leading to anomalous scaling
of the height-height correlation function.
Second, we study a random diffusion model in which one can 
tune the values of the global exponents $\chi$, $\beta$. We compare the exact
solution of the model with numerical integrations of the equation of motion. 
By analyzing the structure factor, we will see that
the second model constitutes an excellent example of what we have termed 
intrinsic anomalous scaling in the previous section, independent 
of the value of $\chi$ below or above one.

\subsection{Linear MBE equation: super-roughening}
Let us start studying the simplest equation relevant for MBE growth 
\cite{villain,linmbe,das1}. For a one dimensional interface it reads
\begin{equation}
{\partial h \over \partial t} = -K {\partial^4 h\over \partial x^4}
+ \eta(x,t),
\label{lmbe}
\end{equation}
where $\eta$ is a white noise with zero mean, and the constant $K>0$. 
Exponents can be easily found by dimensional 
analysis, $\beta=3/8$, $z=4$, and $\chi = 3/2$, thus 
Eq.\ (\ref{lmbe}) generates super-rough surfaces.
This equation has been widely studied by many authors; however for 
the sake of clarity we wish to sumarize here briefly some numerical 
results for it that illustrate our scaling analysis in an
example of super-roughening. 

We have solved numerically Eq.\ (\ref{lmbe}) by means of a finite 
differences scheme.
To study the structure factor, we have calculated $S(k,t)$
in systems of $L=16, 32, 64, 128, 256$. 
\begin{figure}[htb]
\begin{center}
\epsfig{file=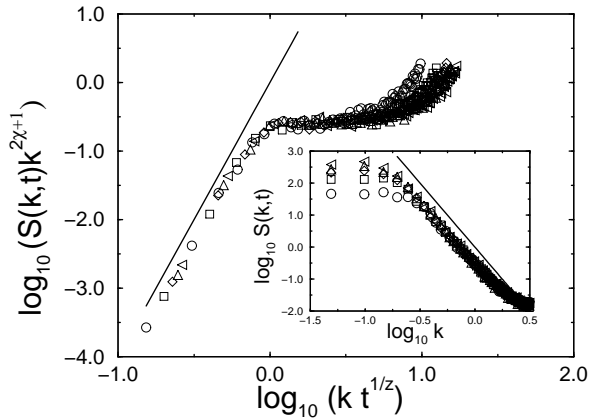, width=2.7in}
\caption{The inset shows the structure factor for the linear MBE equation,
Eq.(21), calculated in a system of total size $L=128$ for times
$t = 10^2, 3\times 10^2, 5\times 10^2, 7\times 10^2, 9\times 10^2 $.
The solid straight line has slope $-4$ and is ploted to guide the eye.
In the main panel the data are collapsed using exponents $\chi = 3/2$, $z=4$.
The solid line has slope 4, showing that the scaling function of the
structure factor is consistent with Eq.\ (10).}
\end{center}
\label{fig1}
\end{figure}
In the inset of Fig.\ 1 numerical results
are shown for $L=128$, and 200 realizations of the noise.
For $k^{-1} \gg t^{1/z}$, the structure factor decays as $k^{-4}$ 
({\em i.\ e.} the global roughness 
exponent $\chi = 3/2$) and does not display any {\it anomaly}.
In Fig.\ 1 we plot a collapse of the data for the exponents
$z=4$ and $\chi=3/2$. As we see, the scaling
function of the structure factor indeed has the Family-Vicsek shape 
(\ref{FVforS}) \cite{curva}. Thus there is no anomaly in the behavior 
of $S(k,t)$ for Eq.\ (\ref{lmbe}).
However, there is anomalous behavior if we look at the local 
properties of the surface in real space. 
We have calculated the local width, $w(l,t)$ for different {\it window} 
sizes $l$ in a system of total size $L=1000$, and averaged over 15 
realizations of the noise. 
\begin{figure}[htb]
\centerline{
\epsfxsize=5cm
\epsfbox{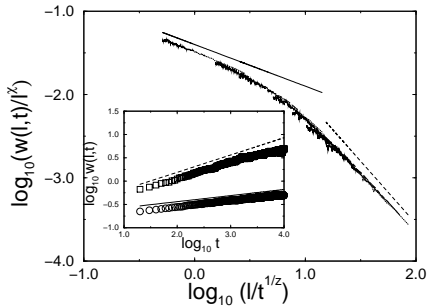}}
\caption{Numerical results for the local width of the linear MBE equation in
a system of size $L=1000$. In the inset we plot $w(l,t)$ vs.\ time for two
different window sizes $l=5$ ($\circ$) and $l=150$ ($\Box$). The straight
lines have slopes 0.13 (solid) and 0.37 (dashed), corresponding
to $\beta_*$ and $\beta$ respectively. Data are collapsed in the main panel for
$l=5$, up to $l=150$ using exponents $\chi = 3/2$, $z=4$. The straight lines
correspond to $u^{-\chi}$ (dashed) and $u^{-(\chi-\chi_{loc})}$ (solid)
and have slopes $-3/2$ and $-1/2$ respectively.}
\end{figure}
An inspection of the inset of Fig.\ 2 shows that the local width on length 
scales $l \ll L$ does not saturate 
at times $t \gg l^z$ but continues to grow in time 
as $w(l,t \gg l^z) \sim t^{\beta_*}$ with $\beta_* \simeq 0.13$, 
in agreement with the theoretical prediction (\ref{Gsuper})
for Eq.\ (\ref{lmbe}) $\beta_* = \beta - 1/z= 0.125$.

As discussed above, in the presence of anomalous scaling the height-height 
correlation function (or the local width) satisfies a dynamic 
scaling form, although with a scaling function which differs from 
the Family-Vicsek one, see Eqs.\ (\ref{FVforG}), (\ref{Anom-G}), and
(\ref{gA}). As stressed in \cite{das2} (for the height-height
correlation function), the plot of 
$w(l,t)/l^{\chi}$ {\it vs.} $l/t^{1/z}$ allows one to determine the form
of the scaling function since 
\begin{equation}
\label{coll}
{w(l,t) \over l^{\chi}} \sim g(l/t^{1/z}).
\end{equation}
The data in the inset of Fig.\ 2 have been collapsed for $l=5, \ldots, 150$,
and plotted in Fig.\ 2. The scaling function $g(u)$ 
obtained is not constant for $u \ll 1$, but scales 
with its argument as $u^{-(\chi-\chi_{loc})}$, with $\chi_{loc}=1$ 
for this case. This is consistent with the scaling behaviour (\ref{Gsuper}), 
or equivalently with (\ref{Anom-G}) for $\chi_{loc}=1$. 

\subsection{Random diffusion model: Intrinsic anomalous scaling }
Next let us consider the more complex case of intrinsic
anomalous scaling. We focus our attention on a growth model with 
a random diffusion coefficient governed by the equation
\begin{equation}
\frac{ \partial h(x,t)}{ \partial t} =
\frac{\partial }{\partial x } \left( D(x) \frac{\partial }{\partial x } 
h(x,t) \right) + \eta (x,t).
\label{columnar}
\end{equation}
The diffusion coefficient $D(x)>0$ is distributed according
to the probability density $P(D) = N_\phi D^{-\phi} f_c(D/D_c)$, 
where the parameter $\phi$ characterizes the
intensity of the disorder. $N_\phi$ is 
merely a normalization constant and the cutoff function 
is $f_c(y) = 1$ for $y \leq 1$ and $f_c(y)=0$ for $y>1$. 
If $\phi < 0$, disorder in the diffusion coefficient does not play any 
role and the Edwards-Wilkinson \cite{ew} exponents $\chi = \chi_{loc} = 1/2$,
$\beta= 1/4$ are recovered. Thus the disorder is termed {\it weak}. 
On the contrary, for {\it strong} disorder, $0 < \phi <1$, the
critical exponents $\chi, \beta$ are disorder-dependent (through the 
value of $\phi$).

There are two major features that make 
this model interesting. On one hand, it is linear and can 
be solved exactly \cite{col,lack}. As a consequence, exponents are 
dependent on the parameter $\phi$ in a known way. In particular, the 
global roughness exponent $\chi$ takes a continuous range of values from
$\chi=1/2$ to $\chi = \infty$ when varying the intensity of disorder from 
$\phi=0$ to $1$. In Ref.\cite{lack} it has been found that the 
random diffusion model always exhibits anomalous scaling for $0 < \phi < 1$. 
Hence anomalous scaling in this model occurs for surfaces with 
roughness exponent $\chi >1$ {\em as well as} for those with $\chi <1$. 
Therefore, this model constitutes an excellent example of anomalous scaling 
not due to super-roughening.

As stated above, the analytical study of (\ref{columnar}) allows to determine 
exactly the critical exponents (see \cite{lack}). For the region of 
interest here, $0<\phi<1$, one has
\begin{equation}
\label{exp}
\beta={1 \over 2(2-\phi)}, \hspace{1cm} \chi={ 1 \over 2(1-\phi)}, 
\hspace{1cm} \chi_{loc} = {1 \over 2} .
\end{equation}
It is a curious 
property of model (\ref{columnar}) that $\chi_{loc}$ remains 
constant when changing the intensity of the disorder, in contrast to 
the wide range of variation existing in $\chi$. This is a feature also found
for many models of rough epitaxial growth \cite{das2}.
In Sec.\ II we have demonstrated that in the case of anomalous kinetic 
roughening the time exponent $\beta_*$ is always given by the 
scaling relation $\beta_*=\beta - \chi_{loc}/z$. 
Thus, from (\ref{exp}), $\beta_*$ depends on $\phi$ as
$\beta_*=\phi/(4-2\phi)$ for a given disorder parameter $\phi$.

We have performed simulations of the discretized version 
of Eq.\ (\ref{columnar}) in a system of lateral size $L=1000$ for 
different degrees of disorder in the strong
disorder phase, $0<\phi<1$. In Fig.\ 3 we plot $w(l,t)$ for two values 
of $\phi$. For $\phi=2/3$, the exact 
global roughness exponent is larger than one, $\chi=3/2$, and one 
observes in the upper panel of Fig.\ 3 that the scaling behaviour 
of the local width is anomalous with $\beta_*=0.27 \pm 0.03$  
in agreement with the theoretical value $\beta_*=1/4$.
\begin{figure}[htb]
\centerline{
\epsfxsize=5cm
\epsfbox{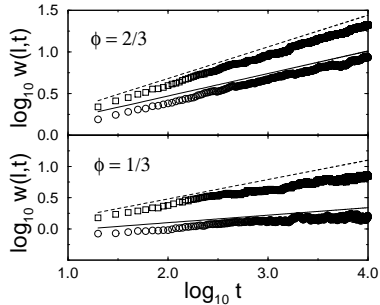}}
\caption{Local width vs.\ time for the random diffusion model, Eq.\ (23),
calculated over windows of size $l=5$ ($\circ$) and $l=150$ ($\Box$) for two
different degrees of disorder. In the upper panel
($\phi = 2/3$), the straight lines have slopes
$0.27$ (solid) and $0.38$ (dashed), to be compared with the exact values
$\beta_* = 1/4$ and $\beta = 3/8$, respectively.
In the lower panel ($\phi=1/3$), the lines have slopes
$0.12$ (solid) and $0.31$ (dashed),  to be compared with
$\beta_*=1/10$ and $\beta=3/10$, respectively.
Thus, the scaling of model (23) is
anomalous ({\em i.e.} $\beta_* \ne 0$) independently of the value of the
global roughenss exponent.}
\end{figure}
In the case of a disorder parameter $\phi =1/3$ (see lower panel of Fig.\ 3), 
for which $\chi=3/4 < 1$, the scaling of $w(l,t)$ is {\em also} anomalous 
with an anomalous time exponent 
$\beta_* = 0.12 \pm 0.04$ to be 
compared to $\beta_*=1/10$ predicted by the exact analysis. 
Thus, in equation (\ref{columnar}) the existence of anomalous 
scaling is completely independent of the surface being super-rough or not. 

As we did above for the linear MBE equation, we 
have determined the scaling function $g(u)$ of the local width
by the data collapse shown in Fig.\ 4, in which the exponents obtained
from Fig.\ 3 have been used.
For both degrees of disorder $\phi = 1/3$ and $2/3$, the scaling function 
behaves as expected for anomalous scaling, and has an analogous shape to that
found for the local width in the super-rough case.
\begin{figure}[htb]
\centerline{
\epsfxsize=5cm
\epsfbox{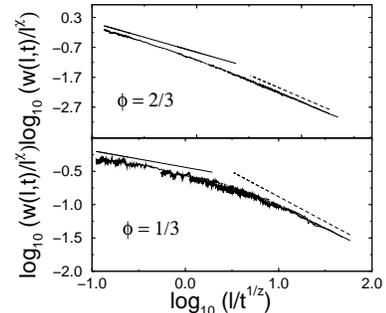}}
\caption{Data collapse of the results displayed in Fig.\ 3. In the upper panel
($\phi= 2/3$), the exponents $\chi = 3/2$, $z=4.08$ have been used. The
straight lines are plotted as a guide to the eye having slopes $-3/2$
(dashed) and $-1$ (solid). For
$\phi = 1/3$ (lower panel), the data show a good collapse for $\chi = 3/4$,
$z=2.42$. The straight lines have slopes $-0.25$ (solid) and $-0.75$
(dashed). In both panels the scaling agrees with formula (25).}
\end{figure}
The complete scaling function $g(u)$ defined in (\ref{coll}) is thus given by
\begin{equation}
g(u) \sim  
\left\{ \begin{array}{lcl}
     u^{-(\chi-\chi_{loc})} & {\rm if} & u \ll 1\\
     u^{-\chi} & {\rm if} & u \gg 1\\
     \end{array}
\right. ,
\end{equation}
where $\chi$ is the disorder-dependent global roughness exponent 
given in (\ref{exp}) and $\chi_{loc}=1/2$.

The {\em intrisic} anomalous character of the scaling in this model 
can be diagnosed through the use of the structure factor. 
We have calculated $S(k,t)$ in systems of sizes $L=16,\ldots,512$. 
Figure 5 shows our results for $L=128$, and 200 realizations of the
disorder. 
\begin{figure}[htb]
\centerline{
\epsfxsize=5cm
\epsfbox{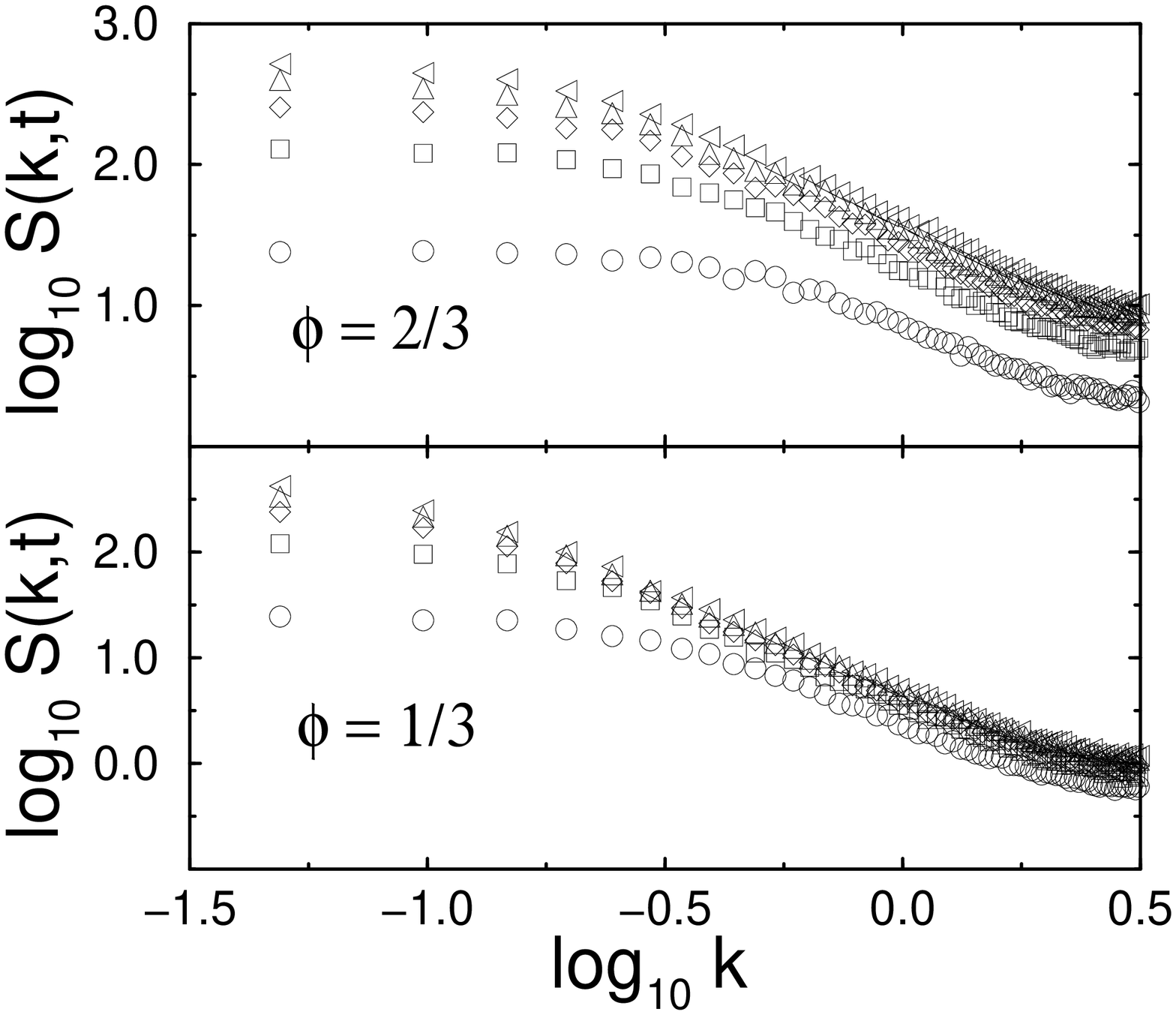}}
\caption{Structure factor of the random diffusion model Eq.\ (23)
for two different degrees of the disorder and times
$t = 10^2, 3\times 10^2, 5\times 10^2, 7\times 10^2, 9\times 10^2 $.
Upper panel, results for $\phi = 2/3$ ($\chi = 3/2$).
Lower panel, data for $\phi = 1/3$ ($\chi= 3/4$). The shift in time
reflects the intrinsic anomalous character of the scaling.}
\end{figure}
For both values of the disorder parameter $\phi=1/3, 2/3$ the spectrum decays 
as $k^{-(2\chi_{loc}+1)}$ (not as $k^{-(2\chi+1)}$) and is
clearly shifted for different times. This scaling behaviour is the one 
in Eq.\ (\ref{Scaling-S}), which we associated in Sec.\ II.B 
with that of growth models having an intrisic anomaly. 
This can be better appreciated when collapsing the curves of Fig.\ 5 as 
shown in Fig.\ 6, which displays the $S(k,t)$ data collapses 
for $\phi = 1/3, 2/3$, and yields a scaling 
function $s(u)$ with a form consistent with Eq.\ (\ref{Anom-S}), and not with
the Family-Vicsek form, Eq.\ (\ref{FVforS}) \cite{curva}. 
\begin{figure}[htb]
\centerline{
\epsfxsize=5cm
\epsfbox{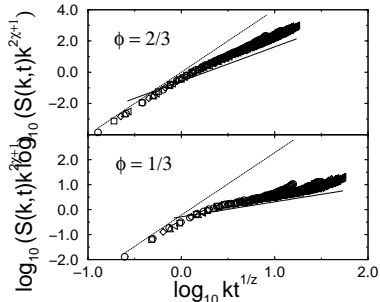}}
\caption{Data collapse of the graphs in Fig.\ 5. The exponents used for the
collapse in the upper panel ($\phi=2/3$) are $\chi=3/2$ and $z=4.08$. The
straight lines have slopes 2.2 (solid) and 4 (dashed). In the lower panel
($\phi=1/3$) the exponents used for the collapse are $\chi=3/4$ and $z=2.42$.
The straight lines have slopes $0.6$ (solid) and $2.5$ (dashed).
In both panels the scaling function is described by Eq.\ (15).}
\end{figure}
As we see, 
this example clearly shows that intrinsic anomalous behaviour may occur 
in the case of a roughness exponent $\chi=3/2 >1$ ($\phi = 2/3$) 
as well as for $\chi = 3/4 < 1$ ($\phi = 1/3$).
 
\section{Conclusions}

As we have seen in the previous sections, one of the main conclusions of the 
present work is that anomalous scaling of rough surfaces is a more general
(and independent) phenomenon than those cases associated with super-roughening.
In this sense several issues associated with anomalus scaling remain to
be clarified. One of them is its relation and/or interplay with the 
phenomena of multifractality and intermitency, known to take place 
in the super-rough case \cite{krug,sg}. Moreover, we have seen that the 
intrinsic anomalous surfaces display a novel type of spectrum thus far not
considered for self-affine geometries, which are usually defined as to have
a Family-Vicsek spectrum \cite{falco}. An outstanding issue in this regard
is to clarify the physical meaning of the exponent $\theta$, as well as 
extending the currently available renormalization group techniques to be able
to calculate the value of this exponent. On the phenomenological side, 
when performing experiments or interpreting numerical simulations of discrete
growth models, we believe the existence of intrinsic anomalous scaling is
a very relevant issue. On one hand, in the presence of intrinsic anomalous
scaling the various correlation functions behave somewhat similarly to the 
Family-Vicsek case. However, indeed there are important differences. For
instance, $S(k,t)$ {\em scales} as $k^{-(2 \chi_{loc}+1)}$, and {\em not}
as $k^{-(2\chi+1)}$, the graphs of $w(l,t)$ and $S(k,t)$ are shifted for
different times, etc. There do exist data collapses, but again they are 
different from the usual Family-Vicsek type. For instance, if
instead of plotting $w(l,t)/l^{\chi}$ vs.\ $l/t^{1/z}$ as recommended in
\cite{das2} and done here in Figs.\ 2, 4, one chooses to represent 
$w(l,t)/t^{\beta}$ vs.\ $l/t^{1/z}$ as is frequently done, one gets, in the
Family-Vicsek case,
\begin{equation}
\frac{w(l,t)}{t^{\beta}} \sim \left\{ \begin{array}{lcl}
{\rm const.} & {\rm if} & l/t^{1/z} \ll 1  \\
\left(\frac{l}{t^{1/z}}\right)^{\chi} & {\rm if} & l/t^{1/z} \gg 1 
\end{array}
\right. ,
\nonumber
\end{equation}
whereas for intrinsic anomalous scaling one gets
\begin{equation}
\frac{w(l,t)}{t^{\beta}} \sim \left\{ \begin{array}{lcl}
{\rm const.} & {\rm if} & l/t^{1/z} \ll 1  \\
\left(\frac{l}{t^{1/z}}\right)^{\chi-\theta} = 
\left(\frac{l}{t^{1/z}}\right)^{\chi_{loc}} & {\rm if} & l/t^{1/z} \gg 1 
\end{array}
\right. ,
\nonumber
\end{equation}
that is, in both cases there is data collapse with a very similar shape 
of the scaling function. However, the slopes of the corresponding scaling
functions for large arguments are different. In the Family-Vicsek case,
the slope coincides with the value of the exponent assumed to achieve the
collapse of the data. However, in the anomalous case it does {\em not}.
Therefore, it is crucial to check whether the slope of the scaling function
does or not coincide with the assumed exponents. We believe this has not
always been done when analyzing data from experiments and/or numerical 
simulations, and may have added to certain confusion existing in the
literature in the identification of universality classes. 
Moreover, the intermediate time regime existing in surfaces with anomalous
scaling introduces difficulties in the evaluation of exponents 
through the common use of local measurements such as the local width,
frequently employed in experiments and numerical simulations. This may lead
to the assignment of erroneous effective values (see \cite{lack}) to the
exponents of surfaces hypothesized to behave in the simple Family-Vicsek
fashion, while a more accurate description of their true scaling properties
may come through the use of the anomalous scaling {\em Ansatz}.

\acknowledgments

We thank E.\ Moro and A.\ S\'anchez for discussions and encouragement.
J.\ M.\ L.\ acknowledges the Postdoctoral Pogram of Universidad de 
Cantabria for support at Ins\-ti\-tu\-to de F\'{\i}sica de Cantabria where most 
of this work has been made. R.\ C.\ also acknowledges warm hospitality at 
Ins\-ti\-tu\-to de F\'{\i}sica de Cantabria. This work has been supported
by DGICyT of the Spanish Government under Project Nr. PB93-0054-C02-02.

\appendix

\section{Local scaling properties of surfaces}

In this Appendix we study the scaling properties of the height-height
correlation function $G(l,t)$ from the different behaviors of the
structure factor $S(k,t)$. In all cases, the starting point is the 
general formula
\begin{eqnarray}
G(l,t) & = & \int_{\frac{2\pi}{L}}^{\frac{\pi}{a}}
\left[1 - \cos\left(k l \right) \right]
\; \frac{s(k t^{1/z})}{k^{2\chi +1}} \; dk \nonumber \\
& = & l^{2\chi} \int_{\frac{2\pi l }{L}}^{\frac{\pi l}{a}}
\left[1 - \cos\left( u \right) \right]
\; \frac{s(\rho u)}{u^{2\chi +1}} \; du , \label{GfromS} 
\end{eqnarray}
where $s(\cdot)$ is the corresponding scaling function of the structure factor 
under study, and $\rho = t^{1/z}/l$.

\subsection{Saturation regime, $l \ll L \ll t^{1/z}$.}

In this regime, $\rho \gg 1$, and the smallest possible value of the 
argument of $s(\rho u)$ in (\ref{GfromS}) is also $\rho \pi l /L \sim 
t^{1/z}/L \gg 1$. Therefore we can replace the scaling function $s(\rho u)$
by its behavior at large arguments, to get the following results:

{\em Family-Vicsek with $\chi < 1$:} In this case the scaling function
$s(\rho u) = s_{FV}(\rho u)$ of Eq.\ (\ref{FVforS}). We get
\begin{equation}
G(l,t) \sim l^{2\chi} \int_{0}^{\infty} 
\frac{1 - \cos u}{u^{2\chi+1}} \; du \sim l^{2\chi} , \label{GFV}
\end{equation}
where the limits $L \rightarrow \infty $ and $a \rightarrow 0$ can be 
taken, and the integral yields a finite constant. 

{\em Family-Vicsek with $\chi \ge 1$:} In this case the scaling function
is as above $s(\rho u) = s_{FV}(\rho u)$. However, for $\chi >1$ 
there is an ultraviolet 
divergence in the integral (\ref{GFV}). We can avoid the divergence by 
keeping finite the lower integration limit ({\em i.e.} taking the limits 
$a \rightarrow 0$, $t^{1/z}/L \rightarrow \infty $ for a fixed $L$). 
In this way we have
\begin{eqnarray}
G(l,t) & \sim & l^{2\chi} \int_{2\pi l/L}^{\infty}
\frac{1 - \cos u}{u^{2\chi+1}} \; du \sim l^{2\chi} 
\left(\frac{l}{L}\right)^{2(1-\chi)} \nonumber \\
& = & l^2 \; L^{2(\chi -1)}, \label{GFVsat}
\end{eqnarray}
where we have kept only the most singular term (as can be shown integrating by 
parts in the above formula). As stated in the text, this implies a local 
value for the roughness exponent $\chi_{loc} = 1$.
The case $\chi =1$ is a marginal situation. Following the same steps
as in Eq.\ (\ref{GFVsat}), one obtains 
\begin{equation}
G(l,t) \sim l^{2 \chi} \log L ,
\end{equation}
that is, $\chi_{loc} = \chi = 1$ up to logarithmic corrections. 

{\em Intrinsic anomalous scaling:}
Now the scaling function we have to use in Eq.\ (\ref{GfromS}) is
$s(\rho u) = s_A(\rho u) \sim (\rho u)^{2 \theta}$, see (\ref{Anom-S}). 
Thus
\begin{equation}
G(l,t) = l^{2\chi} \rho^{2\theta} \int_{0}^{\infty} 
\frac{1 - \cos u}{u^{2\chi+1- 2\theta}} \; du . \nonumber 
\end{equation}
The integrals are convergent as long as $\theta \equiv \chi - \chi_{loc} > 0$
and $\chi_{loc} < 1$ (the latter being always fulfilled as shown in 
\cite{lt,rh}). At saturation we thus have 
\begin{equation}
G(l,t) \sim l^{2(\chi - \theta)} L^{2\theta} = l^{2\chi_{loc}} L^{2\theta} ,
\nonumber
\end{equation}
hence the difference between local and global scaling behavior 
(note, however, $G(l\sim L,t) \sim L^{2(\theta + \chi_{loc})} = L^{2 \chi}$). 

\subsection{Early dynamics, $t^{1/z} \ll l \ll L$.}

In this case the minimum value of the argument of the scaling function 
in (\ref{GfromS}) is $\rho \pi l/L \sim t^{1/z}/L \ll 1$, so that in principle 
the integral probes the whole scaling function $s(\rho u)$, for both 
small and large
values of its argument. In this subsection we first consider the case in which,
further, we have $\rho = t^{1/z}/l \ll 1$. Let us perform a change of variables
in expression (\ref{GfromS}):
\begin{equation}
G(l,t) = l^{2\chi} \rho^{2 \chi} \int_{2\pi t^{1/z}/L}^{\pi t^{1/z}/a}
\left[1 - \cos \left(\frac{y}{\rho}\right)\right] 
\frac{s(y)}{y^{2\chi+1}} \; dy .
\label{Gear}
\end{equation}
We want to study the asymptotics of the integral in (\ref{Gear})
in the limit $\rho 
\rightarrow 0$. Using standard techniques (see {\em e.\ g.} \cite{bo}), 
and assuming $\lim_{y \rightarrow 0} s(y) < \infty$, it is easy to show that,
for $\rho \rightarrow 0$, 
\begin{equation}
\int_{2\pi t^{1/z}/L}^{\pi t^{1/z}/a}
\left[1 - \cos \left(\frac{y}{\rho}\right)\right] \sim 
\int_{0}^{\infty} \frac{s(y)}{y^{2\chi+1}} \; dy + {\cal O}(\rho) , \nonumber
\end{equation}
where the integral appearing in the last equation is well behaved in its 
limits, and $s(y)$ is {\em any} of the scaling functions we are considering,
either $s_{FV}(y)$ or $s_A(y)$. Therefore we get 
\begin{equation}
G(l,t) \sim l^{2\chi} \rho^{2\chi} + {\cal O}(\rho^{2\chi+1})
\sim t^{2\chi/z} \;\;\; {\rm for} \;\;\; t^{1/z} \ll l , \nonumber 
\end{equation}
that is, the scaling properties of the height-height correlation function
in the early dynamics are the same irrespective of the scaling behavior
of the structure factor, whether Family-Vicsek or intrinsically anomalous. 

\subsection{Intermediate time regime, $l \ll t^{1/z} \ll L$.}

Finally, we study the case in which, analogously to the previous subsection, 
the minimum value for the argument of the scaling function along the 
integration region in (\ref{GfromS}) is small ---so
that the whole scaling function contributes in principle--- but the 
parameter $\rho = t^{1/z}/l$ is large. That is, we are in the 
intermediate time regime. In this case, again we have to distinguish between
FV and intrinsic anomalous scaling of $S(k,t)$.

{\em Family-Vicsek scaling with $\chi < 1$:} 
When $s(\rho u) = s_{FV}(\rho u)$ in Eq.\ (\ref{GfromS}), 
the limits $L \rightarrow \infty $, $a \rightarrow 0$ can be taken
after $\rho \rightarrow \infty$, and
\begin{equation}
G(l,t) \sim l^{2\chi} \int_{0}^{\infty} 
(1 - \cos u) \frac{1}{u^{2\chi+1}} \; du \sim l^{2\chi} ,
\label{GinterFV}
\end{equation}
being the integral a finite constant. 

{\em Family-Vicsek scaling with $\chi \geq 1$:}
Again for the case of super-roughening with FV scaling in $S(k,t)$ 
an expansion such as that in Eq.\ (\ref{GinterFV}) leads to a divergent 
integral. The behavior of $G(l,t)$ is best analyzed in the form (\ref{Gear}).
For $\chi>1$, we can take the limits $t^{1/z}/a \rightarrow \infty$,
$t^{1/z}/L \rightarrow 0$ to get
\begin{equation}
G(l,t) \sim l^{2\chi} \rho^{2\chi-2} \int_{0}^{\infty}
\frac{s_{FV}(y)}{u^{y\chi+1}} \sim l^2 \; t^{2(\chi -1)/z} . \nonumber
\end{equation}
In the marginal case $\chi = 1$, the integral in the above expression
diverges when taking the mentioned limits $t^{1/z}/a \rightarrow \infty$,
$t^{1/z}/L \rightarrow 0$, being the most divergent term logarithmic in
time, that is:
\begin{equation}
G(l,t) \sim l^2 \log t .
\nonumber
\end{equation}
Comparing with Eq.\ (\ref{Gsuper}), this yields $\beta_* = 0$ for 
$\chi=1$, meaning logarithmic corrections in time.

{\em Intrinsic anomalous scaling.}
When we use $s(\rho u) = s_A(\rho u)$ in (\ref{GfromS}) and proceed as in
(\ref{GinterFV}),
the behavior of the scaling function at large arguments now contributes with
a leading term (in the limit $\rho \rightarrow \infty$ for fixed
$\rho \; l/a$ and $\rho \; l/L$) of the form
\begin{eqnarray}
G(l,t) & \sim & l^{2\chi} \rho^{2\theta} 
\int_{0}^{\infty} (1 - \cos u) \frac{1}{u^{2\chi+1}} \; du \sim 
l^{2(\chi - \theta)} t^{2\theta/z} \nonumber \\
& = & l^{2\chi_{loc}} t^{2\beta_*} ,
\nonumber 
\end{eqnarray}
hence the anomalous behavior of the height-height correlation function
summarized in (\ref{Anom-G}).

\end{multicols}
\end{document}